\documentclass[prd,reprint,nofootinbib,showpacs,superscriptaddress]{revtex4-1}
\usepackage{graphicx} 
\usepackage{hyperref}
\usepackage{amsfonts}
\usepackage{amsmath,amssymb}
\usepackage{bm} 
\usepackage{color}
\usepackage{epstopdf} 
\usepackage{epsfig}
\usepackage{subfig} 
\usepackage{float}

{\rm }

\def\be{\begin{equation}}
 \def\ee{\end{equation}}
 \def\bea{\begin{eqnarray}}
 \def\eea{\end{eqnarray}}
 \def\bes{\begin{eqnarray}}
 \def\ees{\end{eqnarray}}
 \def\bi{\begin{itemize}}
 \def\ei{\end{itemize}} 

\def\2{\frac{1}{2}}
\def\4{\frac{1}{4}}

\begin{document}

\title{The BB84 quantum key distribution using conjugate homodyne detection}

\author{Bing Qi}
\email{qib1@ornl.gov}
\affiliation{Quantum Information Science Group, Computational Sciences and Engineering Division, Oak Ridge National Laboratory, Oak Ridge, TN 37831, USA}
\affiliation{Department of Physics and Astronomy, The University of Tennessee, Knoxville, TN 37996, USA}

\date{\today}

\begin{abstract}

{Optical homodyne detection has been widely used in continuous-variable (CV) quantum information processing for measuring field quadrature. In this paper we explore the possibility of operating a conjugate homodyne detection system in ``photon counting'' mode to implement discrete-variable (DV) quantum key distribution (QKD). A conjugate homodyne detection system, which consists of a beam splitter followed by two optical homodyne detectors, can simultaneously measure a pair of conjugate quadratures {$X$ and $P$} of the incoming quantum state. In classical electrodynamics, $X^2+P^2$ is proportional to the energy (the photon number) of the input light. In quantum optics, $X$ and $P$ do not commute and thus the above photon-number measurement is intrinsically noisy. This implies that a blind application of standard security proofs of QKD could result pessimistic performance. We overcome this obstacle by taking advantage of two special features of the proposed detection scheme. First, the fundamental detection noise associated with vacuum fluctuations cannot be manipulated by an external adversary. Second, the ability to reconstruct the photon number distribution at the receiver's end can place additional constraints on possible attacks from the adversary. As an example, we study the security of the BB84 QKD using conjugate homodyne detection and evaluate its performance through numerical simulations. This study may open the door to a new family of QKD protocols, in complementary to the well-established DV-QKD based on single-photon detection and CV-QKD based on coherent detection.}

\end{abstract}

\maketitle

\section{Introduction}
\label{sec:1}

Quantum key distribution (QKD) has drawn great attention for the potential to revolutionize cryptography \cite{Gisin02, Scarani09, Lo14, Diamanti16, Xu2020, Pirandola2019}. Presently, the two most well-established families of QKD protocols are discrete-variable (DV) QKD using single-photon detection \cite {BB84, E91} and continuous-variable (CV) QKD using coherent detection (optical homodyne detection) \cite{Ralph99, Hillery00, GMCSQKD}. For simplicity, in this paper we refer them as DV-QKD and CV-QKD correspondingly.

On one hand, DV-QKD protocols, such as the celebrated BB84 QKD \cite {BB84}, have been demonstrated over longer distances \cite{Boaron2018, Liao2017}, and enjoy the more mature security proofs especially when system imperfections and finite data size effects are taken into account. On the other hand, CV-QKD protocols, especially the ones based on coherent states \cite{Diamanti2015}, have shown their own advantages, such as implementable with conventional telecommunication components and potential high key rate at short distances. Note that many distinguishing features of CV-QKD can be contributed to optical homodyne detection, which can be implemented with low-cost photodiodes working at room-temperature. State-of-the-art optical homodyne detectors can be operated above tens of GHz with negligible dead-time and a pathway toward fully integrated, on-chip, photonic implementation \cite{Zhang19}. In addition, the intrinsic filtering provided by the local oscillator in optical homodyne detection can effectively suppress background photons and enable QKD through conventional dense wavelength-division-multiplexed fiber networks in the presence of strong classical traffics \cite{Qi10, Kumar15, Eriksson19} and through daytime free-space channels \cite{Heim2014}. A natural question is: can we implement DV-QKD using optical homodyne detection? If possible, such a hybrid approach may inherit certain advantages from both worlds. In this paper, we address this question by studying the BB84 QKD using conjugate optical homodyne detector operated in ``photon counting'' mode. 

A conjugate homodyne detection system, which consists of a beam splitter followed by two optical homodyne detectors, can simultaneously measure a pair of conjugate quadratures {$X$ and $P$} of the incoming quantum state by maintaining a $90^\textup{o}$ phase offset between the two corresponding local oscillators. In classical electrodynamics, $X^2+P^2$ is proportional to the energy (the photon number) of the input light. In quantum optics, $X$ and $P$ do not commute and thus cannot be determined simultaneously and noiselessly due to Heisenberg’s uncertainty principle. This suggests that the above conjugate homodyne detection is intrinsically noisy. Intuitively, noisy detectors would result poor QKD performance if standard security proofs are applied. To overcome this hurdle, we develop a new security analysis technique exploring two special features of the proposed detection scheme. First, the fundamental detection noise associated with vacuum fluctuations cannot be manipulated by an external adversary (Eve), so it is not necessary to contribute the detector noise to Eve's attack when we estimate an upper bound of Eve's information. This is in line with the trusted detector noise model in CV-QKD \cite{GMCSQKD, Usenko16}. Second, the proposed detection scheme allows the legitimate receiver to reconstruct the photon number distribution of the received light and thus can place additional constraints on the possible attacks from Eve. This is similar to the detector decoy QKD protocol, where the photon number statistics at the receiver's end can be used to improve the performance of QKD \cite{Moroder2009}. As we will show later, by utilizing these two features, a tighter bound on Eve's information can be obtained and an improved secure-key rate can be achieved. We remark that unlike CV-QKD based on phase-sensitive coherent detection, where sophisticated carrier phase recovery scheme may be required to establish a common phase reference between the transmitter (Alice) and the receiver (Bob) \cite{Qi15, Soh15}, the detection scheme proposed in this paper is intrinsically phase insensitive and no phase reference is required. 

This paper is organized as follows: in Sec. II, we review the theory of photon counting using conjugate homodyne detection \cite{Grice1996, Qi2020}, and present two possible ways of applying this detection scheme in the BB84 QKD. In Sec. III, we develop a new security analysis method taking into account the special features of the proposed detection scheme, and conduct numerical simulations to evaluate  the secure-key rates. Finally, in Sec. IV we discuss some practical issues.

\section{The BB84 QKD using conjugate homodyne detection}
\label{sec:2}

\subsection{Conjugate homodyne detection in photon counting mode}

Characterizing photon number statistics using conjugate homodyne detection was investigated in \cite{Qi2020} and the relevant results are summarized in this subsection. The basic setup of a conjugate homodyne detection system is shown in Fig. 1. An unknown quantum state is input from port 1 of a symmetric beam splitter ($\textrm{BS}_1$ in Fig. 1) and a vacuum state is coupled to the other input port. Two optical homodyne detectors are employed to measure the field quadratures of the two output beams of the beam splitter. The phase difference between the two corresponding local oscillators is fixed at $90^\textup{o}$. In this paper, we assume all the optical homodyne detectors are perfect and the input quantum state is in the same mode as the local oscillators. 

\begin{figure}[h!]
\centering\includegraphics[width=8.5cm]{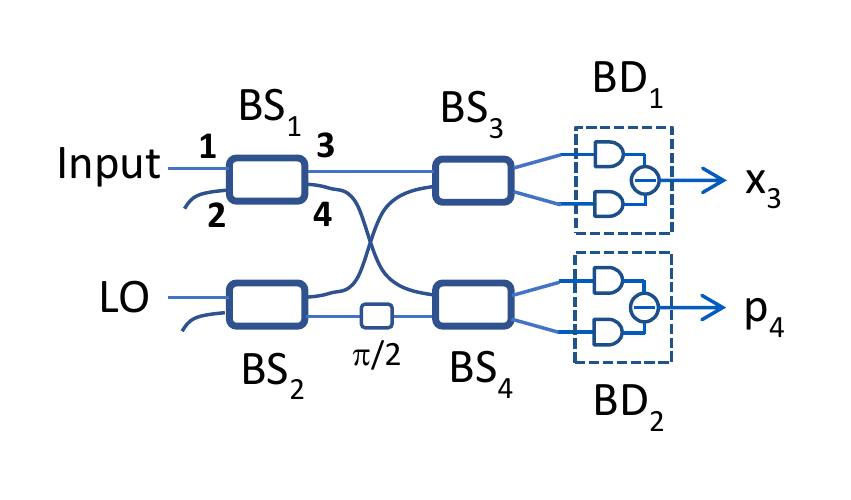}
\caption{Conjugate optical homodyne detection. $\textup{BS}_{1-4}$, symmetric beam splitter; $\textup{BD}_{1-2}$, balanced photodetector; LO, local oscillator.}
\label{fig:1}
\end{figure}

Given the local oscillators are sufficiently strong, the two detection outputs ($X_3$ and $P_4$) are quadrature components of beam 3 and 4 (see Fig. 1). In \cite{Qi2020}, an observable $Z=X_{3}^{2}+P_{4}^{2}$ is defined. Given the density matrix $\rho$ of the input state, the probability density function of $Z$ is given by \cite{Qi2020}
\begin{equation}
P_Z(z) = e^{-z} \sum_{n=0}^\infty \dfrac{\rho_{nn}}{n!}z^n, 
\end{equation}
where $\rho_{nn}$ are the diagonal terms of $\rho$ in Fock basis and $z\geq 0$.

Note that $P_Z(z)$ only depends on the diagonal terms of $\rho$, as expected from a ``phase-insensitive'' photon detector. By repeating the Z-measurement on a large ensemble of identical states, the photon number distribution $P_n=\rho_{nn}$ of the input state can be reconstructed from experimentally determined $P_Z(z)$, as shown in \cite{Qi2020}. This feature allows us to improve the performance of QKD, as we will show in Sec. III.

In the case of single-shot measurement, given the input state is a Fock state $\vert n \rangle$, the likelihood of a measurement output of $z$ can be determined from Eq. (1) \cite{Qi2020}
\begin{equation}
P_Z(z\vert n) = e^{-z} \dfrac{z^n}{n!}.
\end{equation}

In Fig. 2, we present $P_Z(z\vert n)$ for the cases of $n=0,1,2,3$. The overlaps between the probability distributions for different $n$ suggest that the proposed detection scheme is intrinsically noisy. 

\begin{figure}[h!]
\centering\includegraphics[width=8.5cm]{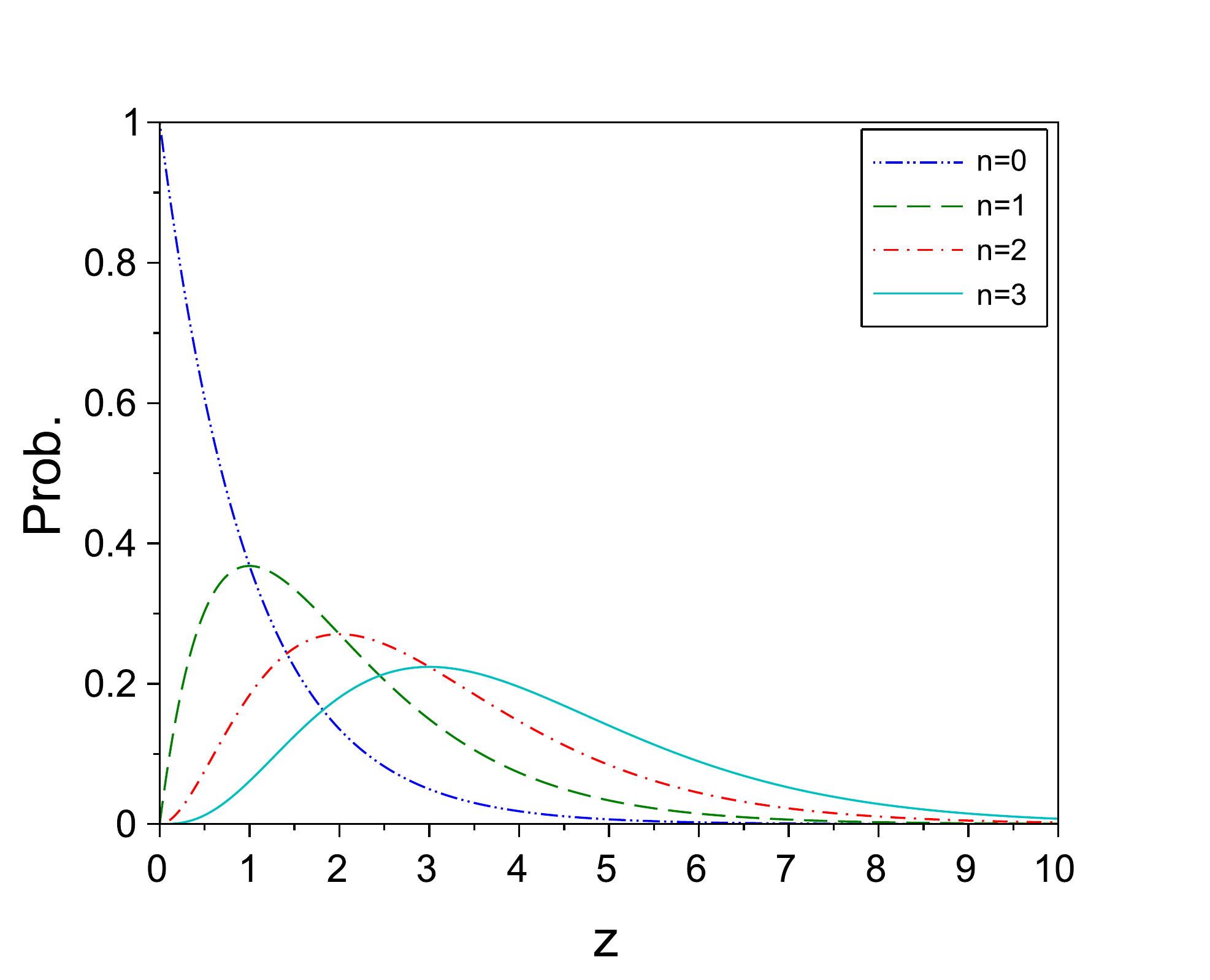}
\caption{(Simulation results) The probability distributions $P_Z(z\vert n)$ corresponding to different input photon number $n$.}
\label{fig:2}
\end{figure}

In practice, single-photon avalanche diodes are commonly used in DV-QKD protocols. This type of single photon detector (SPD) can discriminate vacuum state from non-vacuum states but cannot resolve photon numbers. Its performance can be quantified by single-photon detection efficiency $\eta_D$ and dark count probability $\upsilon_D$, which are defined as the conditional probabilities that the detector clicks given the input is single-photon state or vacuum state, correspondingly.

To operate the conjugate homodyne detector in photon counting mode, we need to map the continuous measurement result $z$ to one of the two possible detection events $\lbrace$click, no-click$\rbrace$. Here, we adopt the same strategy as in \cite{Qi2020}: if $z$ is larger (smaller) than a pre-defined detection threshold $\tau\in\left[0,\infty\right)$, the detector output is assigned as click (non-click). We remark that the above mapping process can be implemented in software in post-processing stage using an optimal $\tau$ adapted to the specific application.

Using Eq. (2), the detection efficiency $\eta_D$ and the dark count probability $\upsilon_D$ can be determined by
\begin{equation}
\eta_D = \int_{\tau}^{\infty} P_Z(z\vert 1) dz = e^{-\tau} (\tau+1)
\end{equation}
\begin{equation}
\upsilon_D = \int_{\tau}^{\infty} P_Z(z\vert 0) dz = e^{-\tau}.
\end{equation}

In Fig. 3, we present $\eta_D$ and $\upsilon_D$ as functions of the detection threshold $\tau$. By choosing an appropriate $\tau$, we could achieve either a high detection efficiency or a low dark count probability, but not both at the same time. In Fig. 3, we also present the ratio $R=\eta_D / \upsilon_D$, which is an important figure of merit in applications like QKD. From Eqs. (3) and (4), $R=\tau+1$, which grows linearly with $\tau$. Unfortunately, $\eta_D$ drops much faster when $\tau$ increases. As shown in Fig. 3, the R-value of the proposed scheme is less than 10 in the region where the detection efficiency is not too low. In comparison, a state-of-the-art SPD can provide a R-value as high as $10^8$ \cite{Hadfield2009}.

\begin{figure}[h!]
\centering\includegraphics[width=8.5cm]{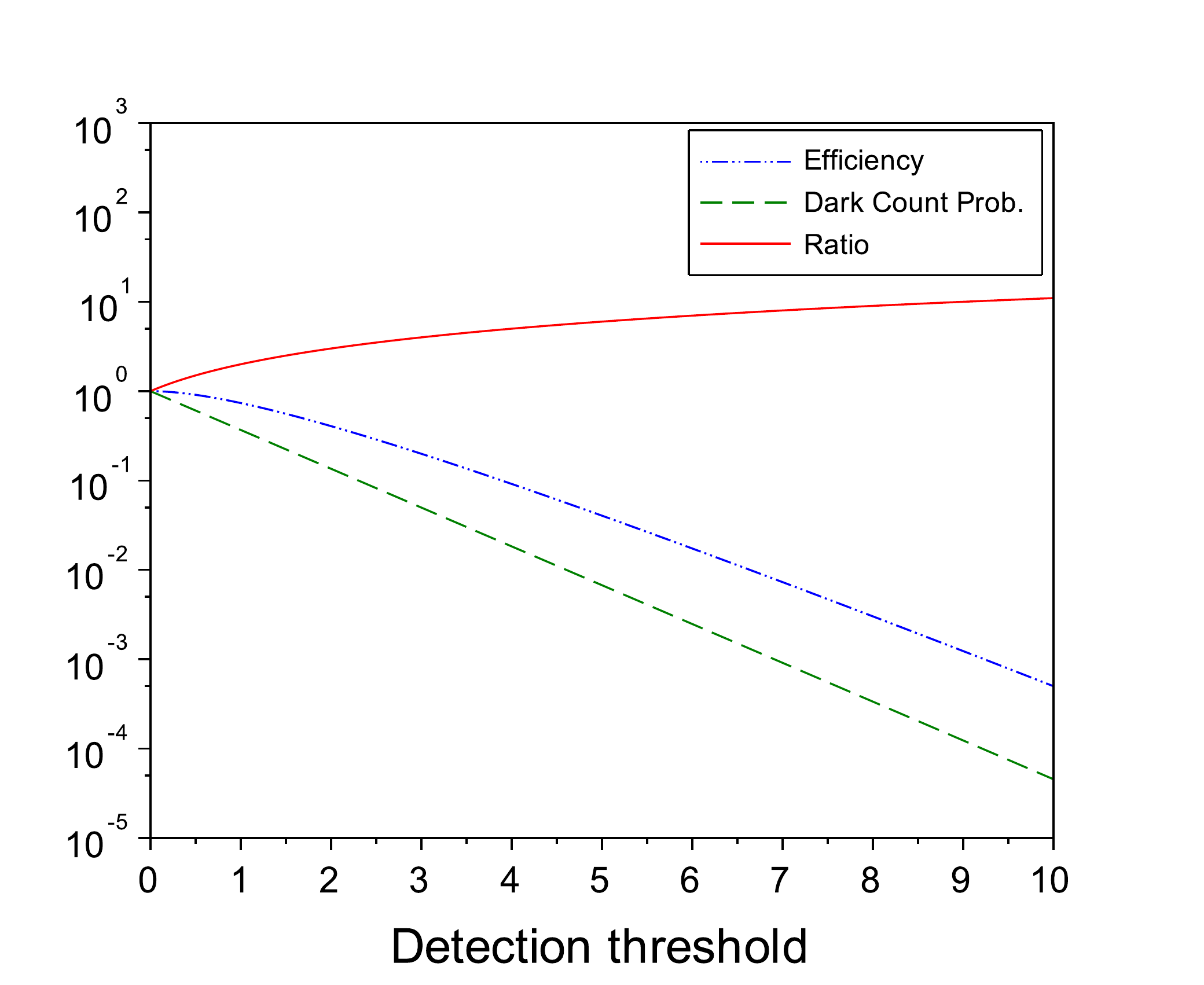}
\caption{(Simulation results \cite{Qi2020}) Detection efficiency $\eta_D$, dark count probability $\upsilon_D$, and the ratio $R=\eta_D / \upsilon_D$.}
\label{fig:3}
\end{figure}

At first sight, the inferior performance of conjugate homodyne detection in photon counting mode seems limit its applications in single-photon based QKD, such as the BB84 protocol. This is probably true if we apply standard security proofs where all the detection noises are contributed to Eve's attack. Such a conservative assumption is necessary when the origins of the noises cannot be identified. In the proposed scheme, the measurement uncertainty of the conjugate homodyne detector is due to fundamental quantum noise rather than technical imperfections. In this case, it is not necessary to contribute the detector noise to Eve's attack. In Sec. III, we present an improved security analysis taking into account the special features of the proposed detection scheme. Below we first summarize the BB84 QKD protocol and the assumptions used in this paper, then discuss two possible ways of using conjugate homodyne detectors in the BB84 QKD.

\subsection{The BB84 QKD and assumptions}

In this paper, we consider the polarization encoding BB84 QKD protocol \cite{BB84}, which includes a quantum stage and a classical post-processing stage. In the quantum stage, for each transmission, Alice prepares a single-photon state with a polarization randomly chosen from $\lbrace$H, V, D, A$\rbrace$, and sends it to Bob through an insecure quantum channel. Here H (V) refers to horizontal (vertical) polarization state and represents bit 0 (1) in the rectilinear basis, while D (A) represents $45^\textup{o}$ ($135^\textup{o}$) polarization state and represents bit 0 (1) in the diagonal basis. At Bob's end, he randomly switches between the two measurement bases using a polarization rotator, measures the polarization of the incoming photons in the chosen basis using a polarizing beam splitter and two detectors ($\textup{D}_0$ and $\textup{D}_1$), and then determines a bit value based on the detection results. In the classical post-processing stage, Alice and Bob use data collected in the diagonal basis to estimate Eve's information and data collected in the rectilinear basis for secure key generation.

We conduct simulations to determine the secure-key rates for both the proposed scheme and the conventional scheme using SPDs. For simplicity, we make the following assumptions through out this paper:
\begin{enumerate}
\item We consider the asymptotic case and neglect any finite data size effects.
\item We adopt the efficient BB84 QKD protocol \cite{LCA2005}, where Alice and Bob chose one basis more often than the other. In the asymptotic case the probability of choosing the preferred basis (the rectilinear basis) approaches to one. 
\item We assume perfect single photon source is employed by Alice. 
\item We assume Bob's detectors are perfect. More specifically, when conventional SPDs are employed, we assume the detection efficiency is one and the dark count probability is zero. When the proposed conjugate homodyne detectors are employed, we assume the quantum efficiency of photo-diodes is one and electrical noises are negligible in comparison to the vacuum noise.
\item We assume perfect error correcting code approaching the Shannon limit is adopted.
\end{enumerate}

Since two conjugate homodyne detectors are required, we consider two possible ways to determine a bit value from the measurement results: independent detection mode and differential detection mode.

\subsection{Independent detection mode}

In this case, $\textup{D}_0$ and $\textup{D}_1$ are operated as two independent photon detectors, with detection efficiency and dark count probability given by Eqs. (3) and (4). This detection mode has been discussed previously in QKD using conventional SPDs \cite{Wang2016}. Here we extend the idea to conjugate homodyne detectors.

Due to the symmetry of the protocol, we assume both detectors use the same detection threshold $\tau$. There are four possible detection outputs \cite{Wang2016}: both detectors click (double-click), only the correct detector clicks, only the wrong detector clicks, and none of them click. The corresponding probabilities are represented by $P_{D}$, $P_{C}$, $P_{W}$, and $P_{N}$.

To give a rough estimation of the potential key rate, we calculate the mutual information $I_{AB}$ under the assumption that there is no technical imperfections except the channel loss. Note $I_{AB}$ is not the secure-key rate, since we do not consider information could be gained by Eve. Nevertheless, it can serve as a rough upper bound on the secure-key rate. We will study lower bounds of secure-key rate in Sec. III.

Given Alice's single photon is prepared in an ideal polarization state corresponding to bit 0, the probability that $\textup{D}_0$ clicks is
\begin{equation}
\begin{split}
&P_{D0}^{(0)} = \eta_{ch}\int_{\tau}^{\infty} P_Z(z\vert 1) dz + (1-\eta_{ch})\int_{\tau}^{\infty} P_Z(z\vert 0) dz\\
&=(\eta_{ch}\tau+1)e^{-\tau},
\end{split}
\end{equation}
where $\eta_{ch}$ is the channel transmittance. 

The probability that $\textup{D}_1$ clicks is simply the dark count probability given by Eq. (4)
\begin{equation}
P_{D1}^{(0)}=e^{-\tau}.
\end{equation}

Since the dark count of $\textup{D}_1$ is independent of the output of $\textup{D}_0$, the probabilities of the four detection events can be determined from Eqs (5) and (6) as
\begin{equation}
P_{N}= (1-P_{D0}^{(0)})(1-P_{D1}^{(0)})=[1-(\eta_{ch}\tau+1) e^{-\tau}] (1-e^{-\tau})
\end{equation}
\begin{equation}
P_{C}= P_{D0}^{(0)}(1-P_{D1}^{(0)})=(\eta_{ch}\tau+1) e^{-\tau} (1-e^{-\tau})
\end{equation}
\begin{equation}
P_{W}= (1-P_{D0}^{(0)})P_{D1}^{(0)}= [1-(\eta_{ch}\tau+1) e^{-\tau}]e^{-\tau}
\end{equation}
\begin{equation}
P_{D}= P_{D0}^{(0)}P_{D1}^{(0)}=(\eta_{ch}\tau+1) e^{-2\tau}.
\end{equation}

We assume that Bob post-selects the single photon detection events and throws away no-click and double-click events. The corresponding gain $Q$ and quantum bit error rate (QBER) $E$ are given by
\begin{equation}
Q=P_{C}+P_{W} = (\eta_{ch}\tau+2) e^{-\tau}-2(\eta_{ch}\tau+1) e^{-2\tau}
\end{equation}
\begin{equation}
E=\dfrac{P_{W}}{Q}=\dfrac{e^{-\tau}-(\eta_{ch}\tau+1) e^{-2\tau}}{Q}.
\end{equation}

The mutual information between Alice and Bob is given by
\begin{equation}
I_{AB}=Q[1-H_2(E)],
\end{equation}
where $H_2(x)=-xlog_2(x)-(1-x)log_2(1-x)$ is the Shannon entropy.

In this paper we assume the quantum channel is standard optical fiber with an attenuation coefficient of $\gamma=0.2dB/km$. The channel transmittance is given by
\bes\label{b2} \eta_{ch}=10^{\frac{-\gamma L}{10}},\ees
where $L$ is the fiber length in kilometers.

Figure 4 (green dashed line) shows the simulation results of $I_{AB}$ as a function of fiber length. In this simulation, the detection threshold $\tau$ is optimized at each distance by maximizing $I_{AB}$.

\begin{figure}[t]
	\includegraphics[width=.45\textwidth]{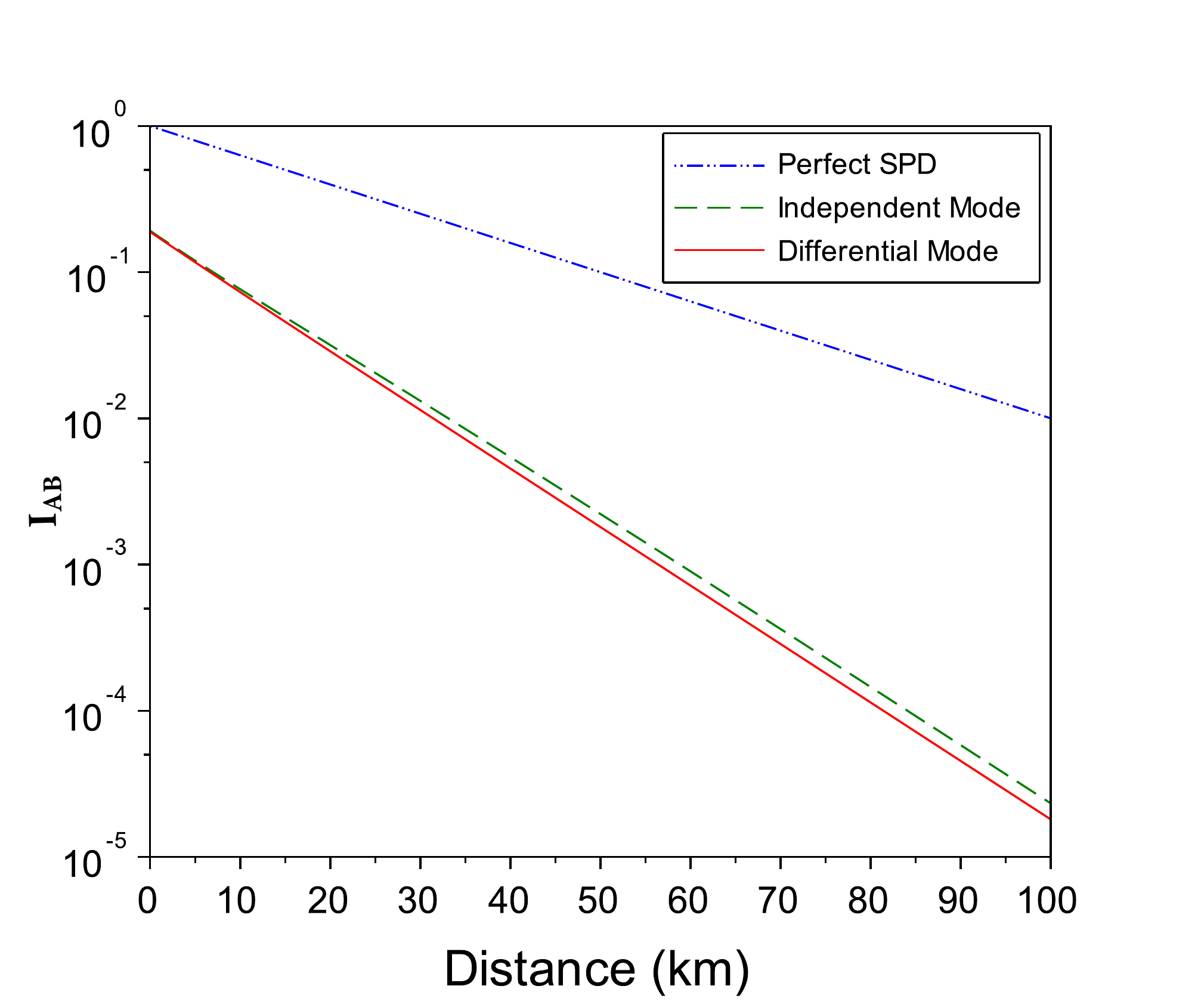}
	\captionsetup{justification=raggedright,
					singlelinecheck=false }
	\caption{The mutual information $I_{AB}$ in three different cases: perfect SPDs, independent detection mode, and differential detection mode. We assume the quantum channel is standard telecom fiber with an attenuation coefficient of $\gamma=0.2dB/km$.}
	\label{fig:4}
\end{figure}

\subsection{Differential detection mode}

In this case, instead of mapping the continuous outputs $z_0$ and $z_1$ of $\textup{D}_0$ and $\textup{D}_1$ into binary detection events independently, we use them jointly to determined the bit value. More specifically, Bob assigns the bit value to 0 (1) if $z_0>z_1$ ($z_1>z_0$). Here, we assume that the probability of $z_0=z_1$ is negligible. Note in this detection mode, there in no need to apply a detection threshold $\tau$. Furthermore, Bob acquires a bit value for every transmission regardless the channel loss, so the gain $Q$ is one.

Given Alice sends bit 0 and the channel transmittance is $\eta_{ch}$, with a probability of $1-\eta_{ch}$, both detectors receive vacuum state. In this case, the error probability (i.e., the probability that $z_0<z_1$) is simply $1/2$. With a probability of $\eta_{ch}$, $\textup{D}_0$ receives one photon and $\textup{D}_1$ receives vacuum. In this case, the error probability is given by $\int_{0}^{\infty} P(z_0\vert 1) \lbrace \int_{z_0}^{\infty} P_Z(z_1\vert 0) dz_1\rbrace dz_0=1/4$, so the average error rate is
\begin{equation}
E=\eta_{ch} \dfrac{1}{4} + (1-\eta_{ch}) \dfrac{1}{2} =\dfrac{1}{2}-\dfrac{\eta_{ch}}{4}.
\end{equation}

Again, $I_{AB}$ can be calculated from Eq. (13) and the result is shown in Fig. 4 (red solid line). As a comparison, in Fig. 4 (blue dash-dotted line), we also present the case when perfect SPDs are employed. In this case, $Q=\eta_{ch}$ and $E=0$, so $I_{AB}=\eta_{ch}$. 

As shown in Fig. 4, $I_{AB}$ determined from the two detection modes are very close to each other. At short distances, both of them are about one order of magnitude below the one achievable with perfect SPDs. Furthermore, schemes using conjugate homodyne detector scale poorer with channel loss than the one using conventional SPDs. While this may look pessimistic at first sight, we remark that an optical homodyne detector could be operated at a much higher detection rate than an SPD. So our proposed scheme could still be a viable solution at short distances. We will discuss this issue more in Sec. IV.

\section{Security analysis}
\label{sec:3}

\subsection{Standard security analysis}
We first calculate the secure-key rate using the standard security proof of the efficient BB84 QKD implemented with a perfect single photon source. The asymptotic secure-key rate is given by \cite{Shor2000}
\begin{equation}
R=Q[1-2H_2(E)],
\end{equation}
where we assume that the QBERs in the two bases are the same.

From Eq. (16), to achieve a positive key rate, $E$ should be less than 11\%. This suggests that the  differential detection mode cannot give a positive key rate, since the minimum error rate is 25\% according to Eq. (15). So we only consider the independent detection mode in this subsection.

Note Eq. (16) is based on the assumption that the quantum state received by Bob is ether vacuum or single-photon state. However, in practice Eve may intercept Alice's photons and send arbitrary quantum state to Bob, so Eq. (16) may not be applied. Fortunately, a detector squashing model exists in the BB84 QKD \cite{Beaudry2008}, which states that as long as the double-click events are kept and assigned with random bit values, Eq. (16) is still applicable. In this case, $Q$ and $E$ can be determined from Eqs. (7)-(10) as
\begin{equation}
Q=1-P_{N} = (\eta_{ch}\tau+2) e^{-\tau}-(\eta_{ch}\tau+1) e^{-2\tau}
\end{equation}
\begin{equation}
E=\dfrac{P_{W}+0.5P_{D}}{Q}=\dfrac{e^{-\tau}-0.5(\eta_{ch}\tau+1) e^{-2\tau}}{Q}.
\end{equation}
We conduct numerical simulations and the asymptotic secure-key rates are shown in Fig. 5 (green dashed line). In this simulation, the detection threshold $\tau$ is optimized at each distance by maximizing the secure-key rate. As a comparison, we also present the secure-key rate for the efficient BB84 using perfect SPDs, which is simply $R=\eta_{ch}$. As shown in Fig. 5, both the secure-key rate and the QKD distance of the new scheme are very limited. To improve the QKD performance, we develop a new security analysis below.

\begin{figure}[t]
	\includegraphics[width=.45\textwidth]{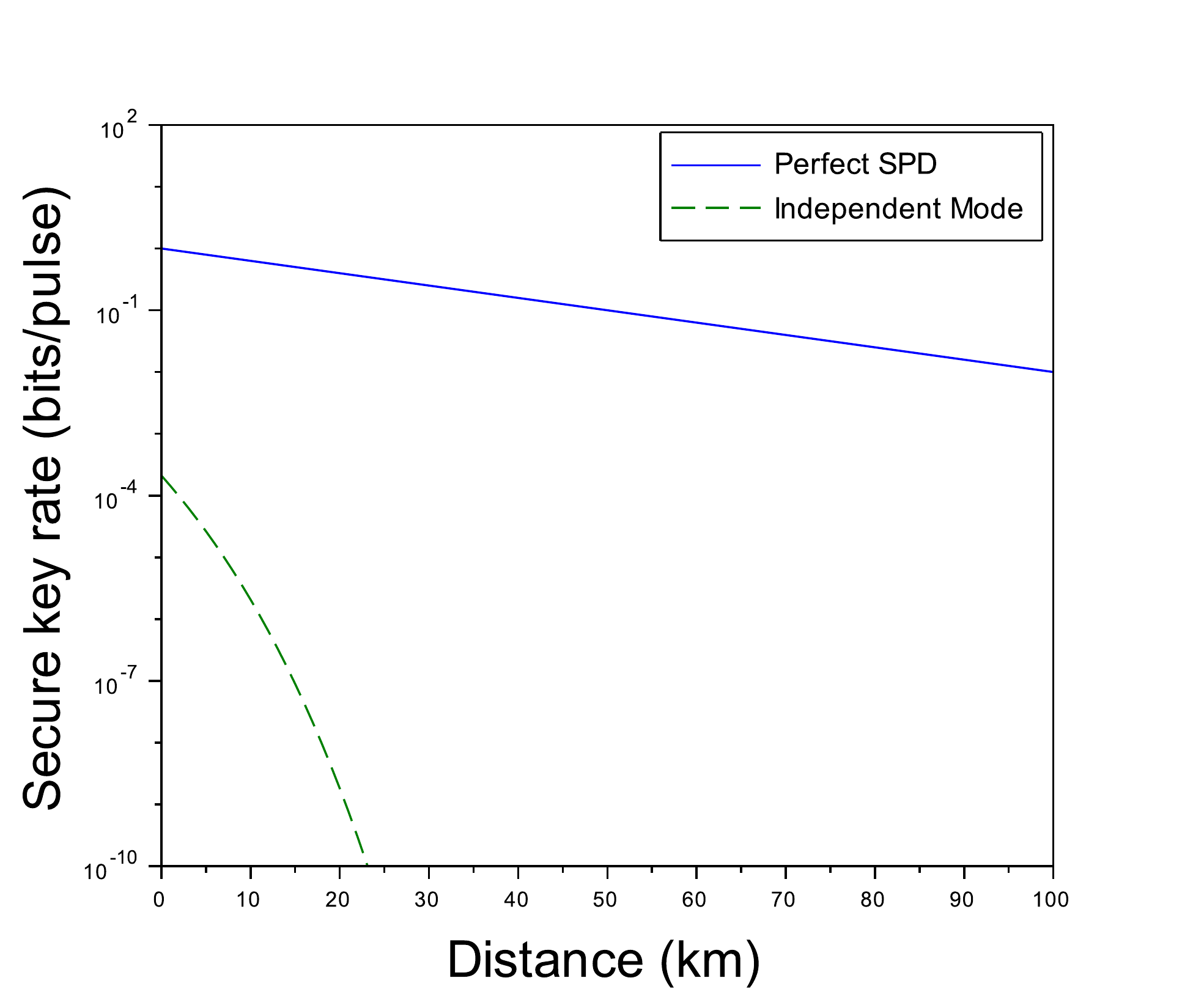}
	\captionsetup{justification=raggedright,
					singlelinecheck=false }
	\caption{Secure-key rates using the standard security proof \cite{Shor2000}. Green dashed line, independent detection mode; Blue solid line, perfect SPDs.} 
	\label{fig:5}
\end{figure}

\subsection{Improved security analysis}

We improve the secure-key rate by taking advantages of two special features of the proposed detection scheme: the quantum origin of the detection noise and the ability of reconstructing the photon number distribution.

We define the joint probability that Alice transmits $m$ photons and Bob receives $n$ photons as $P_{m,n}$, where $m$ and $n$ are nonnegative integers. The corresponding yield (conditional detection probability) and QBER are defined as $Y_{m,n}$ and $E_{m,n}$. Since we assume that a perfect single photon source is employed, the only nonzero terms are $P_{1,n}$, $Y_{1,n}$ and $E_{1,n}$. As we have shown in \cite{Qi2020}, given a large sample size, the photon number distribution of the received quantum state can be reconstructed from the outputs of the detectors. In the asymptotic case, Bob can determine $P_{1,n}$ precisely from his measurement results.

We define the gain of the n-photon state as
\begin{equation}
Q_{1,n}=P_{1,n}Y_{1,n},
\end{equation}
where the n-photon state is defined at Bob's end. Note this is different from the definition in decoy state QKD \cite{Hwang2003, LO2005, Wang2005}, where the n-photon state is commonly defined at Alice's end.

The overall gain $Q$ and the overall QBER $E$ are defined as
\begin{equation}
Q = \sum_{n=0}^\infty Q_{1,n}
\end{equation}
\begin{equation}
E = \frac{\sum_{n=0}^\infty Q_{1,n}E_{1,n}}{Q}.
\end{equation}

Note that in Eqs. (20) and (21), the sum of n goes from zero to infinite, even though Alice transmits only one photon (m=1) to Bob. This is because the quantum channel is controlled by Eve, who may send arbitrary quantum state to Bob. So Bob may receive more than one photons.

To determine the secure-key rate, we consider the reverse reconciliation in the classical post-processing stage \cite{GMCSQKD}, where Bob sends correction information to Alice, who corrects her raw key to have the same values as Bob's. There are three different cases based on the photon number received by Bob:

\textit{Case one: Bob receives a vacuum state}

In this case, both Alice and Eve have no information about Bob's detection results. No secure key can be generated and there is no need to perform privacy amplification. 

\textit{Case two: Bob receives one photon}

In this case, to facilitate the security analysis, we introduce a virtual detection model, as shown in Fig. 6. In this model, a pair of ideal non-demolition SPDs ($\textup{S}_0$ and $\textup{S}_1$), which can determine the photon number without destroying the photons, are placed in front of the real detectors ($\textup{D}_0$ and $\textup{D}_1$). We denote the bit values detected by $\textup{S}_0$ and $\textup{S}_1$ as $\lbrace B^{(V)}_i, i=1,2,...\rbrace$ and the bit values detected by $\textup{D}_0$ and $\textup{D}_1$ as $\lbrace B_i, i=1,2,...\rbrace$.

If Bob could access $\lbrace B^{(V)}_i\rbrace$, then he could generate a secure key from them, and the standard security proof can be applied directly. More specially, the QBER in the rectilinear basis can be used to quantify the cost of error correction, while the QBER in the diagonal basis can be used to upper bound Eve's information thus the cost for privacy amplification. Here we use $E^{(X,V)}_{1,1}$ to represent the QBER in the diagonal basis that could be acquired using the virtual detectors, given Alice sends one photon and Bob receives one photon. Note there is no need to summon to the detector squashing model since Bob receives a qubit. 

\begin{figure}[h!]
\centering\includegraphics[width=8cm]{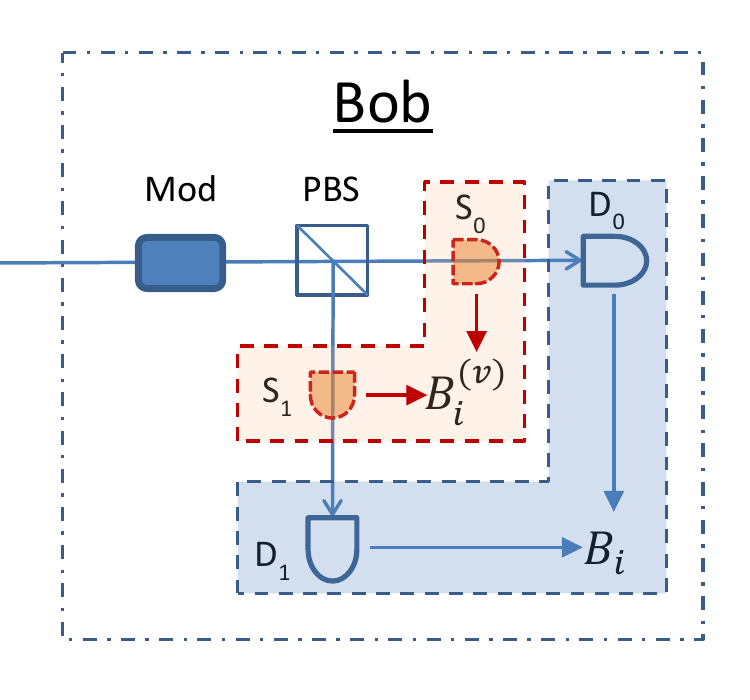}
\caption{A schematic diagram of the virtual detection model. Mod, polarization modulator for basis selection; PBS, polarizing beam splitter;  $\textup{S}_0$ and $\textup{S}_1$, virtual ideal non-demolition SPDs; $\textup{D}_0$ and $\textup{D}_1$, real noisy detectors used by Bob. One key idea is to use the real detectors' outputs $\lbrace B_i\rbrace$ to estimate the quantum bit error rate of the virtual detectors' outputs $\lbrace B^{(V)}_i\rbrace$.}
\label{fig:6}
\end{figure}

In the real protocol,  Bob can only access $\lbrace B_i\rbrace$ measured by the noisy detector $\textup{D}_0$ and $\textup{D}_1$. One important observation is that the detector noise associated with vacuum fluctuations is quantum in nature and cannot be manipulated by Eve. Due to the detector noise, both Alice and Eve's information on $\lbrace B_i\rbrace$ will be no more than their information on $\lbrace B^{(V)}_i\rbrace$. On one hand, the mutual information between Alice and Bob $I_{AB}$ (thus the cost for error correction) can be properly quantified by using the actual QBER measured with the real detectors. On the other hand, as a conservative approach, we can use $E^{(X,V)}_{1,1}$ to quantify Eve's information on $\lbrace B_i\rbrace$ (thus the cost of privacy amplification). We remark that by further quantifying the decrease of Eve's information due to the detector noise, the secure-key rate could be further improved, as we show in Appendix A. 

\textit{Case three: Bob receives more than one photon}

The detector's response to multi-photon signals could be complicated and Eve might be able to introduce basis-dependent detection efficiency by sending tailored multi-photon signals. For simplicity, we assume all the multi-photon signals received by Bob are not secure and cannot be used for secure key generation. Again we remark that by developing a more sophisticated detector model, the secure-key rate could be further improved.

Combined the above three cases, the secure-key rate is given by \cite{LO2005_2}
\begin{equation}
R = Q_{1,0}+Q_{1,1}[1-H_2 (E^{(U,X,V)}_{1,1})]-fQH_2(E),
\end{equation}
where $Q_{1,0}$ represents the contribution from vacuum state, $E^{(U,X,V)}_{1,1}$ represents an upper bound of $E^{(X,V)}_{1,1}$, and $f$ is the error correction efficiency which is assumed to be one in this paper. Below we estimate secure-key rates for the two detection modes. 

\textbf{Secure-key rate: independent detection mode}

To apply Eq. (22) to calculate the secure-key rate, we need to determine five parameters: $Q_{1,0}$, $Q_{1,1}$, $E^{(U,X,V)}_{1,1}$, $Q$ and $E$. Since $Q$ and $E$ can be determined from experimental data directly, below we discuss how to determine the rest.

From Eq. (19), $Q_{1,0}=P_{1,0}Y_{1,0}$ and $Q_{1,1}=P_{1,1}Y_{1,1}$. As we noted early, the photon number distribution $P_{1,n}$ can be reconstructed from Bob's detection results. Furthermore, we do not need to call for the detector squashing model and can simply throw away all the no-click and double-click events. Using Eq. (2), we have
\begin{equation}
\begin{split}
&Y_{1,0} = 2\int_{0}^{\tau} P_Z(z_0\vert 0) dz_0 \times \int_{\tau}^{\infty} P_Z(z_1\vert 0) dz_1\\
&=2(1-e^{-\tau})e^{-\tau}. 
\end{split}
\end{equation}

The corresponding QBER is 
\begin{equation}
E_{1,0} = 0.5.
\end{equation}

Similarly, under the assumption that the two detector $\textup{D}_0$ and $\textup{D}_1$ are identical, $Y_{1,1}$ is independent of the polarization state of the received photon, and can be determined by 
\begin{equation}
\begin{split}
&Y_{1,1} = \int_{0}^{\tau} P_Z(z_0\vert 1) dz_0 \times \int_{\tau}^{\infty} P_Z(z_1\vert 0) dz_1\\
&+\int_{\tau}^{\infty} P_Z(z_0\vert 1) dz_0 \times \int_{0}^{\tau} P_Z(z_1\vert 0) dz_1\\
&=(\tau+2)e^{-\tau}-2(\tau+1)e^{-2\tau}. 
\end{split}
\end{equation}

The corresponding QBER is
\begin{equation}
\begin{split}
&E_{1,1} = (1-E^{(V)}_{1,1}) \frac{\int_{0}^{\tau} P_Z(z_0\vert 1) dz_0 \times \int_{\tau}^{\infty} P_Z(z_1\vert 0) dz_1}{Y_{1,1}}\\
&+E^{(V)}_{1,1}\frac{\int_{0}^{\tau} P_Z(z_0\vert 0) dz_0 \times \int_{\tau}^{\infty} P_Z(z_1\vert 1) dz_1}{Y_{1,1}}\\
&=\frac{(E^{(V)}_{1,1} \tau +1)e^{-\tau}-(\tau+1)e^{-2\tau}}{Y_{1,1}},
\end{split}
\end{equation}
where $E^{(V)}_{1,1}$ is the expected QBER from the virtual ideal non-demolition SPDs.

Using Eq. (21), we have
\begin{equation}
\begin{split}
&QE = Q_{1,0}E_{1,0}+Q_{1,1}E_{1,1}+\sum_{n=2}^\infty Q_{1,n}E_{1,n}\\
&\geq Q_{1,0}E_{1,0}+Q_{1,1}E_{1,1},
\end{split}
\end{equation}
which leads to an upper bound of $E_{1,1}$
\begin{equation}
E_{1,1}\leq E_{1,1}^{(U)}= \frac{QE-Q_{1,0}E_{1,0}}{Q_{1,1}}.
\end{equation}

Note Eqs. (23) to (28) can be applied in both bases. Once an upper bound of $E_{1,1}$ in the diagonal basis has been obtained from Eq. (28), an upper bound of $E^{(X,V)}_{1,1}$ can be determined by using Eq. (26). By now, all the parameters needed in Eq. (22) have been derived.

To evaluate the QKD performance, we calculate the secure-key rate under normal operating conditions without Eve's attack. Since we assume a perfect single photon source is employed, for a pure loss channel, Bob either receives a vacuum state or a single-photon state, with the corresponding probabilities of $P_{1,0}=1-\eta_{ch}$ and $P_{1,1}=\eta_{ch}$. All the other probabilities $P_{1,n}=0$ for $n\geq 2$. We further assume that the QBER due to polarization misalignment is $E_d$. Using the above photon number distribution, it is easy to show that $Q_{1,0}=(1-\eta_{ch})Y_{1,0}$, $Q_{1,1}=\eta_{ch}Y_{1,1}$, $Q=Q_{1,0}+Q_{1,1}$, $E=\frac{0.5Q_{1,0}+Q_{1,1}E_{1,1}}{Q}$, and $E^{(U,X,V)}_{1,1}=E_d$, where $Y_{1,0}$, $Y_{1,1}$, and $E_{1,1}$ are given in Eqs. (23), (25) and (26).

The simulation results are shown in Fig. 7. As a comparison, we also present the secure-key rate of the BB84 QKD implemented with perfect SPDs and no polarization misalignment ($E_d=0$). Comparing with the results shown in Fig. 5, the QKD performance has been greatly improved. Again we optimize the detection threshold $\tau$ at each distance by maximizing the secure-key rate. The corresponding optimal values of $\tau$ are shown in Fig. 8. Note that in the case of $E_d=0.05$, there is a jump of the optimal $\tau$ in Fig. 8 around 8.2 km, which leads to the non-differentiability on the corresponding curve in Fig. 7. We investigate this phenomenon by calculating the secure-key rate as a function of $\tau$ at fixed distances. In the case of $E_d=0.05$, there are two local optimal values of $\tau$. As the distance increases, the global optimal $\tau$ switches from the first local optimum to the second one at the distance around 8.2 km, which results a jump of $\tau$. 

\begin{figure}[t]
	\includegraphics[width=.45\textwidth]{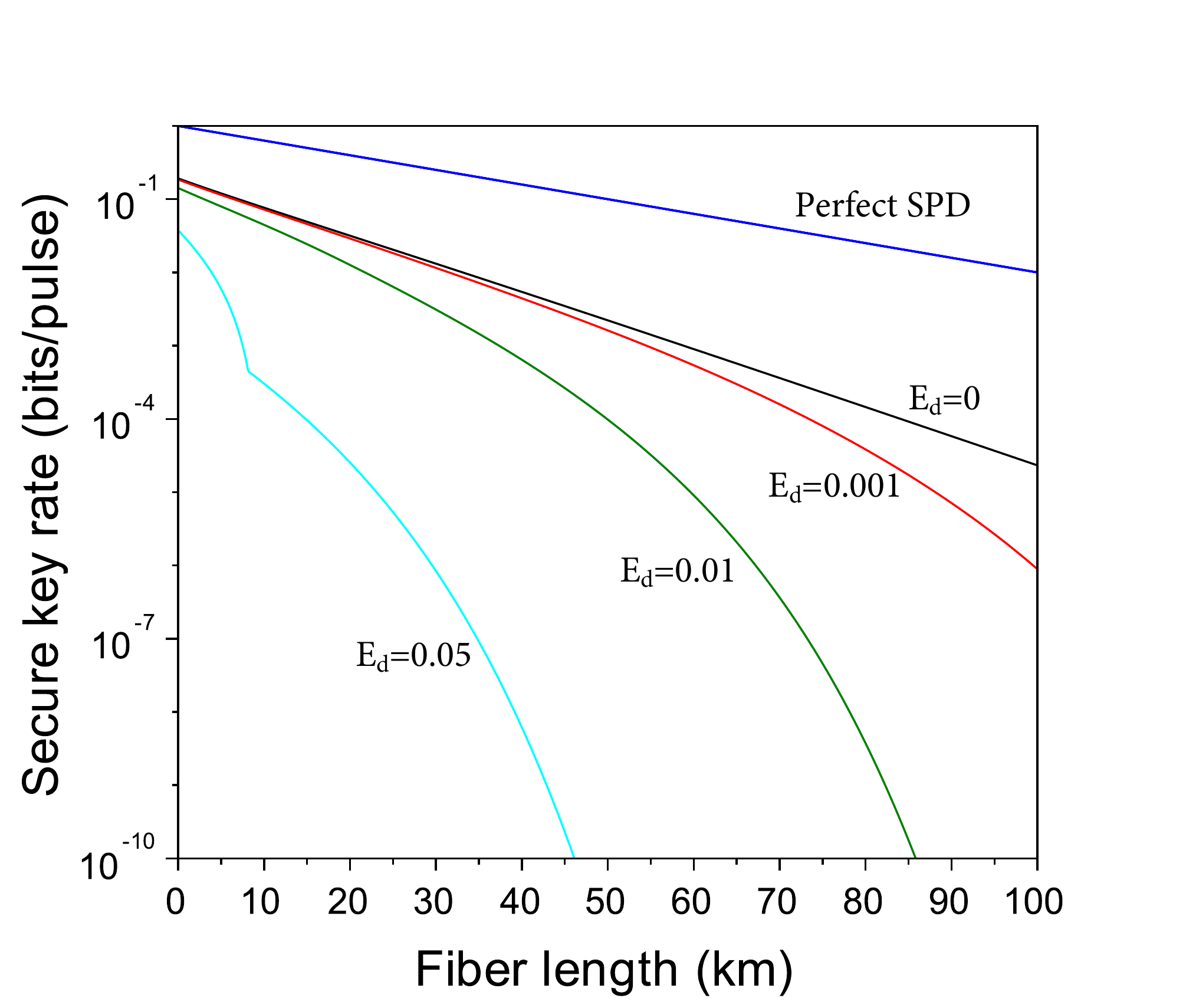}
	\captionsetup{justification=raggedright,
					singlelinecheck=false }
	\caption{Secure-key rates using the improved security analysis (independent detection mode). $E_d$, error probability due to polarization misalignment. As a comparison, the secure-key rate of the BB84 QKD implemented with perfect SPDs at $E_d=0$ is also presented.} 
	\label{fig:7}
\end{figure}

\begin{figure}[t]
	\includegraphics[width=.45\textwidth]{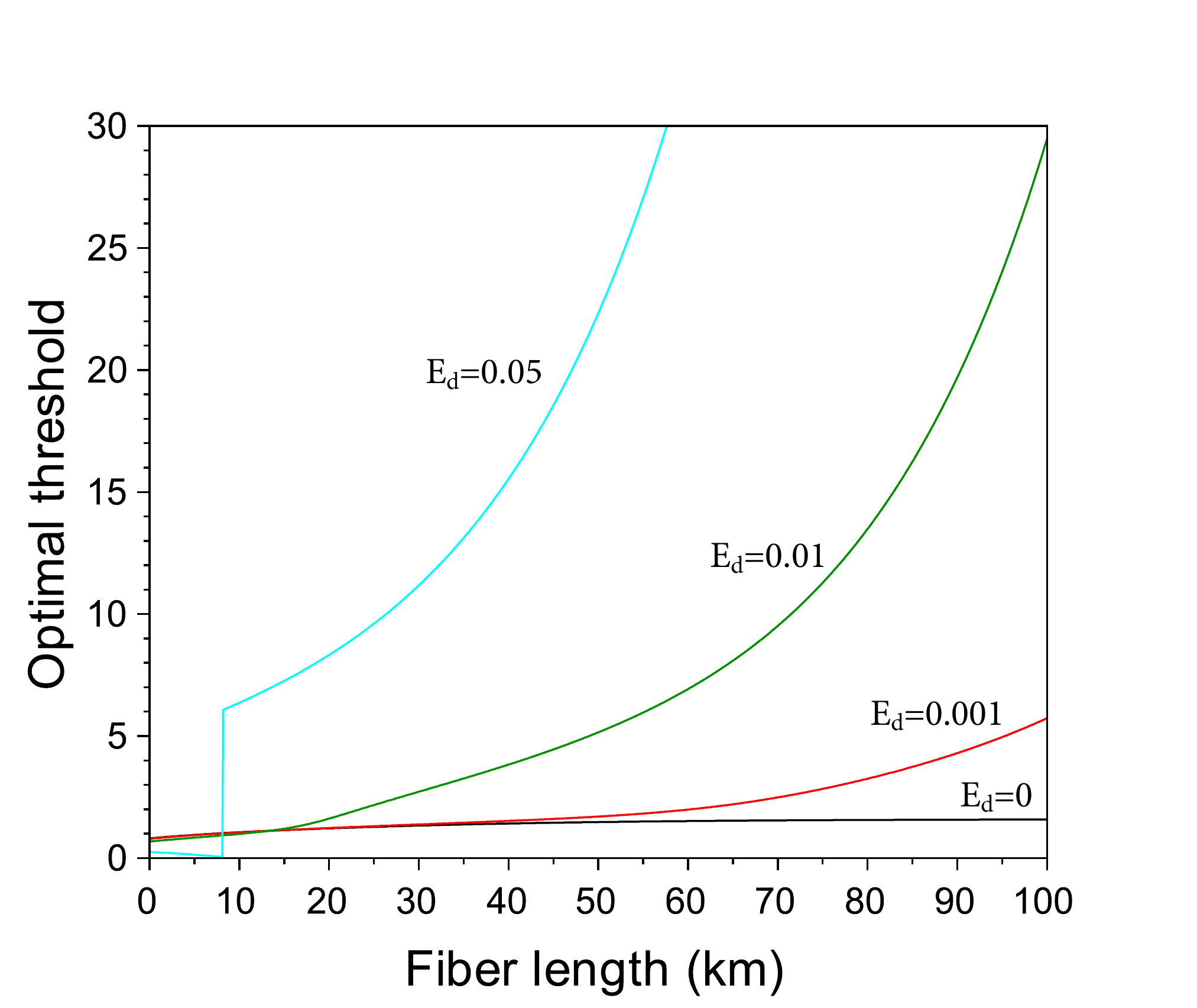}
	\captionsetup{justification=raggedright,
					singlelinecheck=false }
	\caption{Optimal detection threshold $\tau$. Note there is a jump of $\tau$ around 8.2 km when $E_d=0.05$ (see discussion in the main text).} 
	\label{fig:8}
\end{figure}

\textbf{Secure-key rate: differential detection mode}

The analysis in the differential detection mode is similar to that in the independent detection mode but with a few modifications.

First, in this mode, Bob's detectors work in a deterministic fashion, meaning for each transmission Bob's detectors will output either bit 0 or bit 1. This suggests that $Y_{1,n}=1$ for any $n$. So $Q_{1,n}=P_{1,n}$ and $Q=1$. 

Second, $E_{1,0}$ is still 0.5, but $E_{1,1}$ is given by
\begin{equation}
\begin{split}
&E_{1,1} = (1-E^{(V)}_{1,1}) \int_{0}^{\infty} P_Z(z_0\vert 1) \lbrace \int_{z_0}^{\infty} P_Z(z_1\vert 0) dz_1 \rbrace dz_0\\
&+E^{(V)}_{1,1}\int_{0}^{\infty} P_Z(z_0\vert 0) \lbrace \int_{z_0}^{\infty} P_Z(z_1\vert 1) dz_1 \rbrace dz_0\\
&=\frac{1}{4}+\frac{E^{(V)}_{1,1}}{2}.
\end{split}
\end{equation}

Again, once an upper bound of $E_{1,1}$ in the diagonal basis has been obtained from Eq. (28), an upper bound of $E^{(X,V)}_{1,1}$ can be determined from Eq. (29).

The simulation results are shown in Fig. 9. Note no secure key can be generated when $E_d=0.05$.
Comparing with the results shown in Fig. 7, while both detection modes yield similar key rates when $E_d=0$, the independent detection mode can tolerate higher polarization misalignment.

\begin{figure}[t]
	\includegraphics[width=.45\textwidth]{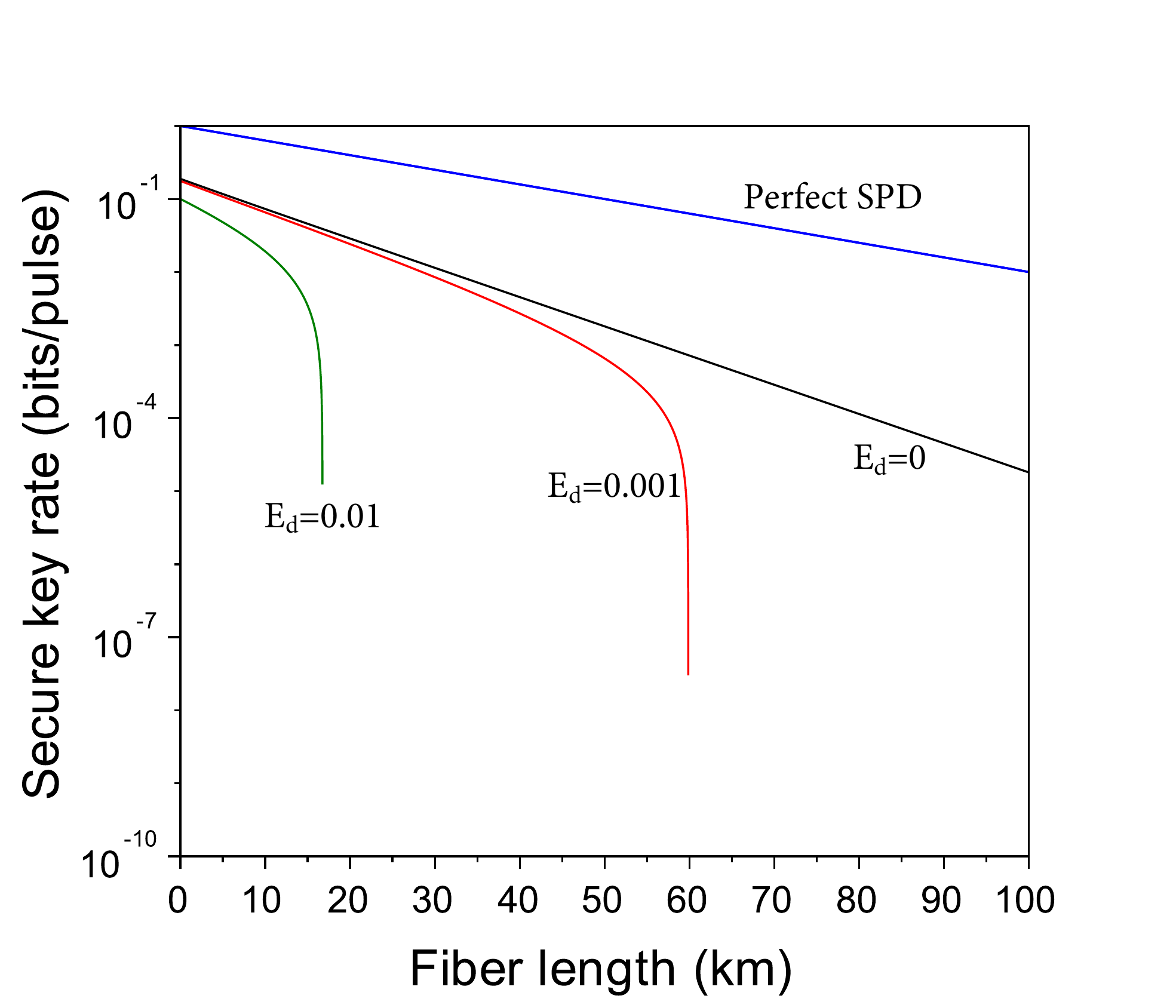}
	\captionsetup{justification=raggedright,
					singlelinecheck=false }
	\caption{Secure-key rates using the improved security analysis (differential detection mode). No secure key can be generated when $E_d=0.05$. As a comparison, the secure-key rates of the BB84 QKD implemented with perfect SPDs at $E_d=0$ are also presented.} 
	\label{fig:9}
\end{figure}

\section{Discussion}
\label{sec:4}

In this paper, we present a hybrid QKD protocol by implementing the BB84 QKD using optical homodyne detectors. Can it be a useful solution in practice? Below we discuss two potential advantages of the proposed scheme, together with detector side-channel attacks and other implementation issues. 

\textit{High secure-key rate over short distances}. The secure-key rates shown in Figs. 7 and 9 are given in bits per transmission. In practice, secure-key rates quantified in bits per second are more relevant. Limited by its dead-time, the maximum detection rate of a practical SPD is typically below 100 MHz. This places a constrain on the achievable secure-key rate of the BB84 QKD implemented with SPDs: regardless how high the transmission rate is, the final secure-key rate cannot be larger than the detection rate. On the contrary, state-of-the-art optical homodyne detectors in classical optical coherent communications can be operated above tens of GHz with negligible dead-time. By further reducing electrical noise, those detectors could be used in the proposed QKD. In fact, the key part of a high-speed optical homodyne detector, shot-noise limited balanced photodiodes with a bandwidth of 5 GHz, is commercially available \cite{HOM}. Equipped with the above high-speed detectors, the new QKD scheme could provide higher secure-key rates (in bits per second) than the conventional scheme over short distances.

\textit{Robust against broadband background photons}. The proposed detection scheme can be implemented with highly efficient photodiodes working at room-temperature and is highly integratable. More importantly, the intrinsic filtering provided by the local oscillator in optical homodyne detection can effectively suppress background photons and enable QKD through conventional dense wavelength-division-multiplexed fiber networks in the presence of strong classical traffics \cite{Qi10, Kumar15, Eriksson19} and through daytime free-space channels \cite{Heim2014}. By removing the requirement of establishing a phase reference between Alice and Bob, the proposed ``phase insensitive'' detection scheme is easier to implement than the ``phase sensitive'' coherent detection scheme used in CV-QKD. To detect Alice's photon efficiently, Bob's local oscillators should be in the same mode as Alice's photon. This requirement is equivalent to generate indistinguishable photons from  two isolated lasers, which has been routinely demonstrated with commercial off-the-shelf lasers in the so-called measurement-device-independent (MDI) QKD \cite{LO2012}.

\textit{Detector side-channel attacks}. QKD protocols are unconditionally secure in theory. However, their real-life implementations can never be perfect. This opens the door to various side-channel attacks. Especially, SPDs in the conventional DV-QKD protocols are regarded as the most vulnerable part for two reasons. First, the quantum channel is controlled by Eve, who can send arbitrary quantum states to Bob's detectors. It is difficult to predict the detector's response to an unknown input state. Second, the extremely high sensitivity of an SPD is a double-edged sword: on one hand, it allows Bob to detect a single photon efficiently. On the other hand, it makes the detector more vulnerable to external attacks. Since an optical homodyne detector is designed to work with both quantum and classical signals, we expect our QKD scheme is more robust against detector side-channel attacks.

One illustrative example is the detector-blinding attack on SPDs. In one implementation \cite{Lydersen2010}, Eve first shines bright light on Bob's detectors to convert them from threshold SPDs to classical linear detectors, then introduces basis-dependent detection efficiency by sending tailored faked states and gain some information of the secure key. With an optical homodyne detector, it could be more difficult to introduce such a dramatic change of the detector property without being detected. In fact, in our detection scheme, Bob is able to reconstruct the photon number distribution of the received signal, and all the multi-photon signals are assumed to be insecure (see Set. IIIB). The bright light from Eve may result a high QBER and expose her presence. Similar argument could also be applied to the saturation attack in CV-QKD \cite{Qin2016}, where Eve displaces the quantum signals to the saturation region of the detector. 

There are also more subtle attacks exploring certain asymmetry of the Bob's detection system. For example, in time-shift attack \cite{Qi2007}, Eve takes advantage of the efficiency mismatch between different SPDs in time domain and gains partial information of Bob's detection results by manipulating the arrival time of the quantum signal at Bob's detectors. Similarly, in wavelength attack \cite{Li2011}, Eve explores the wavelength-dependence of Bob's system and launches the attack by sending lights with carefully chosen wavelengths. The proposed detection scheme could be more resilient to the above attacks, thanks to the intrinsic filtering provided by the local oscillator. By generating the local oscillators for different optical homodyne detectors from a common light source, the detection efficiency of different detectors can be well matched in both time and spectral domain. Since all the local oscillators are generated locally by Bob, the proposed scheme is also immune to many side-channel attacks in CV-QKD using transmitted local oscillator, where Eve launches her attack by manipulating both the quantum signal and the local oscillator \cite{Ma2013, Jouguet2013, Huang2013}.

While the proposed detection scheme may be more resilient against detector side-channel attacks than the conventional SPD, it is impossible to eliminate all the side-channels. In practice, it is still important to investigate potential loopholes and develop corresponding countermeasures. We remark one appealing solution to all detector side-channel attacks is MDI-QKD \cite{LO2012}, where both Alice and Bob transmit quantum signals to an untrusted measurement device, which could be fully controlled by Eve. Unfortunately, the proposed detection scheme may not be applicable in MDI-QKD. This is because in our security analysis, we explicitly assume that the detector noise is trusted and cannot be accessed by Eve. 

\textit{Other implementation issues}. To apply the proposed scheme in practice, there are several challenges to be addressed. First, in the present study we assume a perfect single photon source is employed. In practice, most of the BB84 QKD implementations are based on phase-randomized weak laser sources, which can generate more than one photon occasionally and are susceptible to photon-number splitting attack \cite{Brassard2000}. Fortunately, this problem has been solved in conventional BB84 QKD by introducing the so-called decoy state protocols \cite{Hwang2003, LO2005, Wang2005}, where by randomly modulating the intensity of weak laser pulses, the detection statistics of single-photon states can be acquired. It could be an interesting research topic to incorporate the decoy state idea into our scheme. Second, the technical imperfections of optical homodyne detectors, including the non-unity quantum efficiency and electrical noise, are ignored in this study. Those imperfections need to be quantified and taken into account in the security analysis. Finally, in this paper we only consider asymptotic cases where all the QKD parameters can be determined precisely. It is important to further investigate the case with finite data size.

In summary, we explore the possibility of operating optical homodyne detectors in photon counting mode to implement DV-QKD protocols. By developing a new security analysis based on the special features of the detector, we show that reasonable secure-key rates could be achieved. This study may open the door to a new family of QKD protocols, in complementary to the well-established DV-QKD based on single-photon detection and CV-QKD based on coherent detection.

We acknowledge helpful comments from Nicholas A. Peters and Brian P. Williams. This work was performed at Oak Ridge National Laboratory (ORNL). ORNL is managed by UT-Battelle, LLC, under Contract No. DE-AC05-00OR22725 for the U.S. Department of Energy (DOE). We acknowledge support from the DOE Office of Cybersecurity Energy Security and Emergency Response (CESER) through the Cybersecurity for Energy Delivery Systems (CEDS) program.

The U.S. Government retains and the publisher, by accepting the article for publication, acknowledges that
the U.S. Government retains a nonexclusive paid-up irrevocable worldwide license to publish or reproduce the published form of this manuscript, or allow others to do so, for U.S. Government purposes. The DOE will provide public access to these results of federally sponsored research in accordance with the DOE Public Access Plan \cite{DOE}.

\appendix

\section{A tighter bound on Eve's information}

In Sec. IIIB, we quantify Eve's information on Bob's raw key $\lbrace B_i\rbrace$ using $H_2(E^{(U,X,V)}_{1,1})$, where $E^{(U,X,V)}_{1,1}$ is an upper bound of the QBER in the diagonal basis that would be acquired using the perfect virtual detectors, given Alice sends one photon and Bob receives one photon (see Eq. (22) in the main text). This is a conservative approach, since $H_2(E^{(U,X,V)}_{1,1})$ upper bounds Eve's information on $\lbrace B^{(V)}_i\rbrace$ (the outputs of the virtual detectors). Due to the detector noise, Eve's information on $\lbrace B_i\rbrace$ will be less than her information on $\lbrace B^{(V)}_i\rbrace$. In this appendix, we derive a tighter bound on Eve's information using classical probability theory. This result may be applicable to individual attacks. We leave the case of general attacks for future study.

Below we present the details for the case of independent detection mode. The analysis for the differential detection mode is similar. We denote Eve's estimations of $\lbrace B^{(V)}_i\rbrace$ as $\lbrace E_i\rbrace$. The corresponding bit error rate $E^{(V)}_{EB}$ is defined as
\begin{equation}
E^{(V)}_{EB}=P(B^{(V)}_i=1|E_i=0),
\end{equation}
where we assume $P(B^{(V)}_i=1|E_i=0)=P(B^{(V)}_i=0|E_i=1)$.

Since Eve's information on $\lbrace B^{(V)}_i\rbrace$ is upper bounded by $H_2(E^{(U,X,V)}_{1,1})$, we have
\begin{equation}
I^{(V)}_{EB}=1-H_2(E^{(V)}_{EB}) \leq H_2(E^{(U,X,V)}_{1,1}).
\end{equation}

Similar to Eq. (26) in the main text, the bit error rate between $\lbrace B_i\rbrace$ and $\lbrace E_i\rbrace$ can be determined by
\begin{equation}
\begin{split}
&E_{EB} = P(B_i=1|E_i=0)\\
&= P(B_i=1|B^{(V)}_i=0)\times P(B^{(V)}_i=0|E_i=0)\\
&+P(B_i=1|B^{(V)}_i=1)\times P(B^{(V)}_i=1|E_i=0)\\
&= \frac{\int_{0}^{\tau} P_Z(z_0\vert 1) dz_0  \int_{\tau}^{\infty} P_Z(z_1\vert 0) dz_1}{Y_{1,1}}(1-E^{(V)}_{EB})\\
&+\frac{\int_{0}^{\tau} P_Z(z_0\vert 0) dz_0 \int_{\tau}^{\infty} P_Z(z_1\vert 1) dz_1}{Y_{1,1}}E^{(V)}_{EB}\\
&=\frac{(E^{(V)}_{EB} \tau +1)e^{-\tau}-(\tau+1)e^{-2\tau}}{Y_{1,1}},
\end{split}
\end{equation}
where $Y_{1,1}$ is given by Eq. (25) in the main text.

Once $E^{(U,X,V)}_{1,1}$ has been determined following the steps in the main text, a lower bound of $E^{(V)}_{EB}$ can be determined from Eq. (A2). Consequentially, a lower bound of $E_{EB}$ can be determined from Eq. (A3). We further use $1-H_2(E_{EB})$ as an estimation of Eve's information on $\lbrace B_i\rbrace$ and calculate the secure-key rate by replacing the $H_2(E^{(U,X,V)}_{1,1})$ term in Eq. (22) with $1-H_2(E_{EB})$.

The simulations results for the independent detection mode and the the differential detection mode are shown in Fig. 10 and Fig. 11 correspondingly. Comparing with the results shown in Fig. 7 and Fig. 9, the QKD performance has been improved for the cases of $E_d\neq 0$.

\begin{figure}[t]
	\includegraphics[width=.45\textwidth]{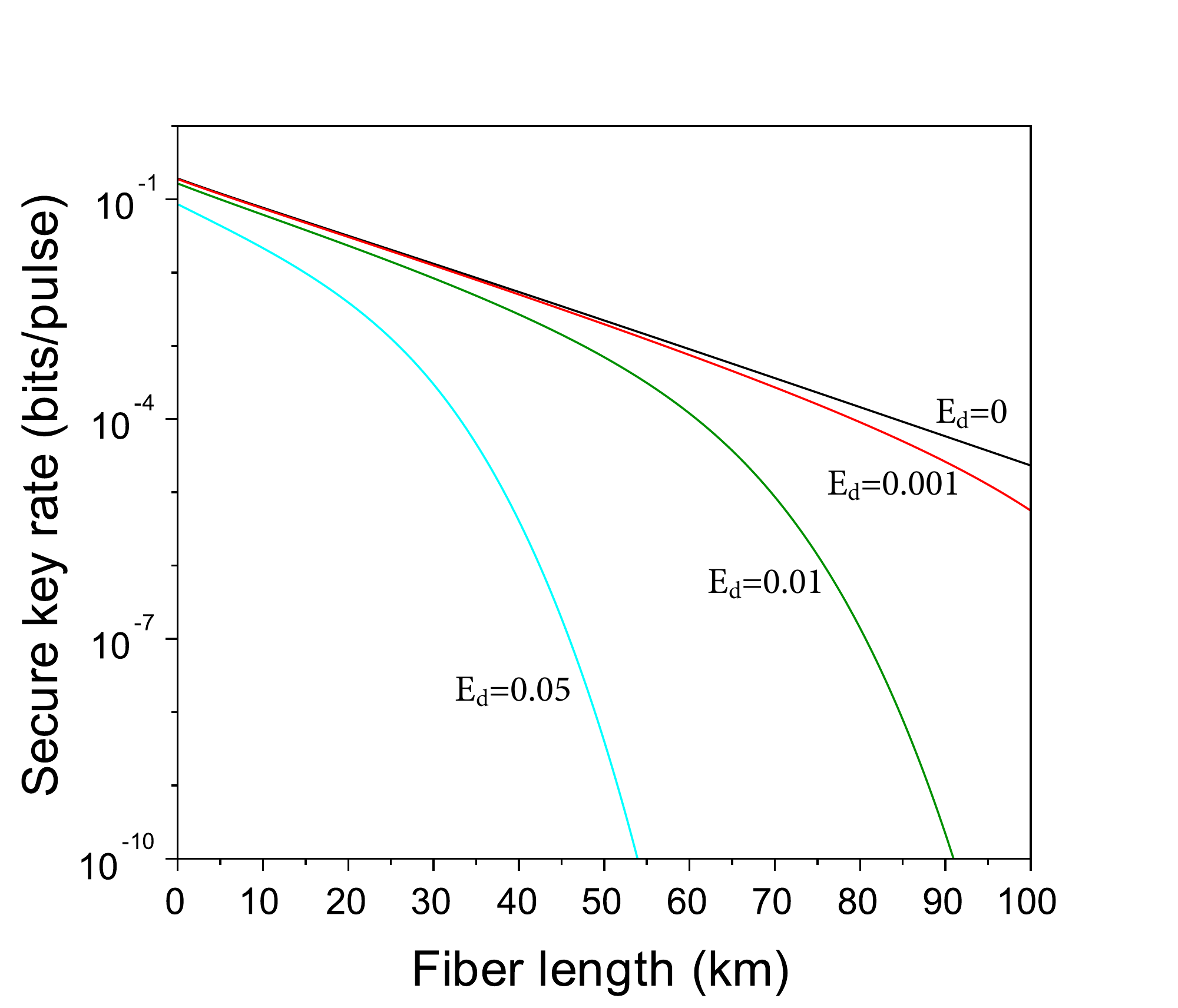}
	\captionsetup{justification=raggedright,
					singlelinecheck=false }
	\caption{Secure-key rates using the tighter bound in Appendix A (independent detection mode).} 
	\label{fig:10}
\end{figure}

\begin{figure}[t]
	\includegraphics[width=.45\textwidth]{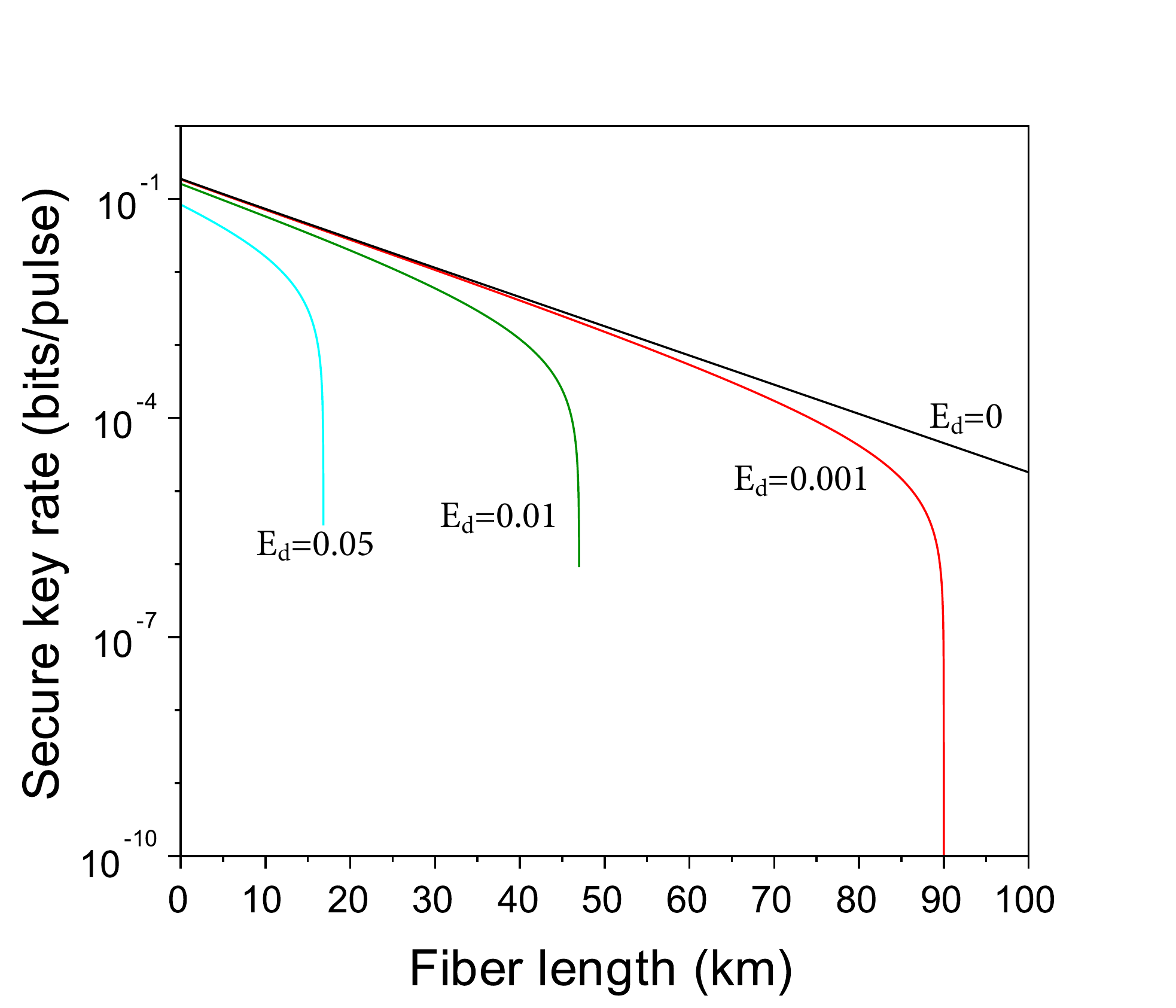}
	\captionsetup{justification=raggedright,
					singlelinecheck=false }
	\caption{Secure-key rates using the tighter bound in Appendix A (differential detection mode).} 
	\label{fig:11}
\end{figure}

\end{document}